\documentclass[amsmath,amssymb,twocolumn,pra,superscriptaddress]{revtex4}
\usepackage{amsfonts}
\usepackage{amsmath}
\usepackage{amsthm}
\usepackage{amscd}
\usepackage{amssymb}
\usepackage{subfigure}
\usepackage{amsxtra}
\usepackage{gensymb}
\usepackage{bm}           
\usepackage{bbm}
\usepackage{graphicx}
\usepackage{epstopdf}
\usepackage{color}
\usepackage{times}
\usepackage{mdframed}
\usepackage[colorlinks=true,citecolor=blue,urlcolor=black]{hyperref}

\def\bi#1\ei {\begin{itemize}#1\end{itemize}}
\def\bn#1\en {\begin{enumerate}#1\end{enumerate}}
\def\bea#1\eea {\begin{align}#1\end{align}}
\def\bean#1\eean {\begin{align*}#1\end{align*}}
\def\ben#1\een {\begin{equation*}#1\end{equation*}}
\def\be#1\ee {\begin{equation}#1\end{equation}}
\def\bes#1\ees {\begin{equation}\begin{split}#1\end{split}\end{equation}}
\def\bear#1\eear {\begin{eqnarray}#1\end{eqnarray}}
\def\bear#1\eear {\begin{eqnarray*}#1\end{eqnarray*}}

\emergencystretch=\maxdimen
\newcommand{\beq}{\begin{equation}}
\newcommand{\eeq}{\end{equation}}
\newcommand{\bra}[1]{\ensuremath{\langle#1|}}
\newcommand{\ket}[1]{\ensuremath{\left|#1\right\rangle}}
\newcommand{\braket}[2]{\ensuremath{\langle #1|#2\rangle}}

\begin{document}

\title{\bf Characterizing multi-photon quantum interference with practical light sources and threshold single-photon detectors}

\author{\'Alvaro Navarrete}
\affiliation{EI Telecomunicaci\'on, Department of Signal Theory and Communications, University of Vigo, Vigo E-36310, Spain}

\author{Wenyuan Wang}
\affiliation{Center for Quantum Information and Quantum Control, Department of Physics and Department of Electrical \& Computer Engineering, University of Toronto, Toronto, Ontario, M5S 3G4, Canada}

\author{Feihu Xu}
\affiliation{National Laboratory for Physical Sciences at Microscale, University of Science and Technology of China (Shanghai Brunch), Hefei, Anhui 230026, P. R. China}

\author{Marcos Curty}
\affiliation{EI Telecomunicaci\'on, Department of Signal Theory and Communications, University of Vigo, Vigo E-36310, Spain}

\date{\today}

\begin{abstract}
The experimental characterization of multi-photon quantum interference effects in optical networks is essential in many applications of photonic quantum technologies, which include quantum computing and quantum communication as two prominent examples. However, such characterization often requires technologies which are beyond our current experimental capabilities, and today's methods suffer from errors due to the use of imperfect sources and photodetectors. In this paper, we introduce a simple experimental technique to characterise multi-photon quantum interference by means of practical laser sources and threshold single-photon detectors. Our technique is based on well-known methods in quantum cryptography which use decoy settings to tightly estimate the statistics provided by perfect devices. As an illustration of its practicality, we use this technique to obtain a tight estimation of both the generalized Hong-Ou-Mandel dip in a beamsplitter with six input photons, as well as the three-photon coincidence probability at the output of a tritter.
\end{abstract}

\maketitle

\section{Introduction}

Multi-photon quantum interference is a key concept in quantum optics and quantum mechanics. It has been extensively studied by many authors over the last decades, going from the seminal two-photon interference experiment performed by Hong, Ou and Mandel~\cite{hong1987measurement} to more recent experimental demonstrations which involve a higher number of indistinguishable photons in various scenarios~\cite{ou1999observation,liu2007demonstration,liu2007four,xiang2006demonstration,niu2009observation,kobayashi2016frequency,lopes2015atomic}. Moreover, besides its indubitable inherent theoretical interest, multi-photon quantum interference also plays a pivotal role on several subfields and applications of quantum information science that use, for example, optical networks (ONs) to interfere photons. These applications include, among others, quantum computing~\cite{knill2001scheme}, quantum cryptography~\cite{lo2012,yin2016}, boson sampling~\cite{aaronson2011computational,broome2013photonic,spring2013boson,tillmann2013experimental,crespi2013integrated}, quantum clock synchronization~\cite{quan2016demonstration}, and quantum metrology~\cite{bell2013multicolor}. In any practical realization of these applications it is essential to experimentally confirm that the photons interfere as desired~\cite{obrien2009photonic}.

Unfortunately, however, to experimentally characterize multi-photon quantum interference on general ONs is usually a quite challenging task~\cite{lobino2008complete}. This is so because, for this, one would ideally need to use high-quality on-demand $n$-photon sources which are yet to be realized~\cite{PhysRevLett.117.210502,PhysRevLett.116.020401}, together with high-quality photon-number resolving (PNR) detectors, which, besides of being expensive experimental resources, currently can only distinguish up to a certain number of photons and may also introduce noise~\cite{kardynal2008,lita2008,harder2016local}. As a result, we have that current experimental techniques to characterize the quantum interference behaviour of ONs at a few photons level typically suffer from inevitable errors due to the use of imperfect sources and detectors~\cite{lobino2008complete}.

The main contribution of this paper is a novel technique to experimentally estimate the input-output photon number statistics of ONs when the input signals are tensor products of Fock states. For this, we use simple laser sources to generate the input signals to the ON and practical threshold single-photon detectors to measure the output signals~\cite{alvaro2015interferencia}. That is, our method is implementable with current technology and allows the estimation of the conditional probability distribution $P(x_1,...,x_M|n_1,...,n_N)$ that describes the behaviour of the ON on the input Fock states $\ket{n_1,...,n_N}$, where $n_i$ ($x_j$), with $i=1,...,N$ ($j=1,...,M$), denotes the number of photons at the $i$th ($j$th) input (output) port of the ON. We emphasize, however, that, in practice, our method is specially suitable to evaluate mainly small-size ONs. This is so because, as we show later, it requires to experimentally estimate the probabilities of certain observable events whose estimation complexity may increase exponentially with the number of input/output ports of the ON~\cite{aaronson2011computational,troyansky1996quantum}.

The key idea builds on two techniques that are extensively used in the field of quantum cryptography: the decoy-state method~\cite{hwang2003,lo2005,wang2005a} and the so-called detector-decoy technique~\cite{moroder2009,lim2015random}. We use the former at the input ports of the ON to estimate the statistics provided by ideal $n$-photon sources. Besides standard quantum key distribution~\cite{hwang2003,lo2005,wang2005a,peng2007,Rosenberg2007,yuan2007}, the decoy-state method has also been used for example to estimate the yield of two single-photon pulses in measurement-device-independent quantum key distribution~\cite{lo2012,curty2014,xu2013}, to simulate single-photons sources with imperfect light sources~\cite{Yuan2016simulating}, and to perform single-photon quantum state tomography with practical sources~\cite{Valente2017probing}. That is, so far the use of the decoy-state method has been limited to evaluate the behaviour of ONs when they receive single-photon pulses at their input ports. Here we extend its use to estimate the behaviour of ONs in the general case where they receive as input signals multi-photon pulses. Furthermore, we employ the detector-decoy method at the output ports of the ON to estimate the statistics provided by ideal PNR detectors~\cite{moroder2009,lim2015random}.

To illustrate the practicality of our technique to study ONs, we evaluate two simple examples of interest. In the first one, we estimate the \textit{generalized} Hong-Ou-Mandel (HOM) dip~\cite{hong1987measurement} in a beamsplitter when the total number of input photons is six for two different conditional probabilities, $P(3,3|3,3)$ and $P(5,1|5,1)$. The first case has been 
experimentally studied in~\cite{xiang2006demonstration}, where the authors used for this a spontaneous parametric down-conversion source in combination with a measurement setup with six threshold single-photon detectors. The second case, however, (to the best of our knowledge) has not been experimentally implemented yet due to the difficulty of generating five-photon states to input the beamsplitter. In both scenarios we use our method to estimate the HOM dip by means of just two laser sources and two threshold single-photon detectors. In the second example, we estimate the three coincidence detection probability in a tritter~\cite{Messen2017distinguishability} when there is just one single-photon pulse in each of its input ports, i.e., we estimate $P(1,1,1|1,1,1)$. This example is also used to obtain a high precision estimation of the dependence of that probability with the {\it triad} phase, which arises when one considers more than two input photons~\cite{Messen2017distinguishability}. While these two examples correspond to evaluating linear ONs, we remark that our method could also be used to study multi-photon quantum interference in non-linear ONs. 

The paper is organized as follows. In Sec.~\ref{SecMethod}, we present our method in detail. Then, in Sec.~\ref{SecEvaluation} we evaluate the two practical examples described above. Finally, we summarize the content of the paper in Sec.~\ref{SecConclusions}. The paper also includes two appendixes with additional information.

\section{Method}\label{SecMethod}
As already mentioned above, we use the decoy-state (detector-decoy) method at the input (output) ports of the ON to estimate the statistics provided by ideal $n$-photon sources (PNR detectors). Of course, in contrast to the case where one really uses perfect $n$-photon sources and PNR detectors, the use of decoy settings does not provide single shot resolution about how many photons input and output each port of the ON each given time. However, it permits to estimate the {\it full statistics} that such perfect devices could give, which is enough for our purposes. 

\begin{figure}
	\begin{center}
		\includegraphics[angle=0,scale=0.34]{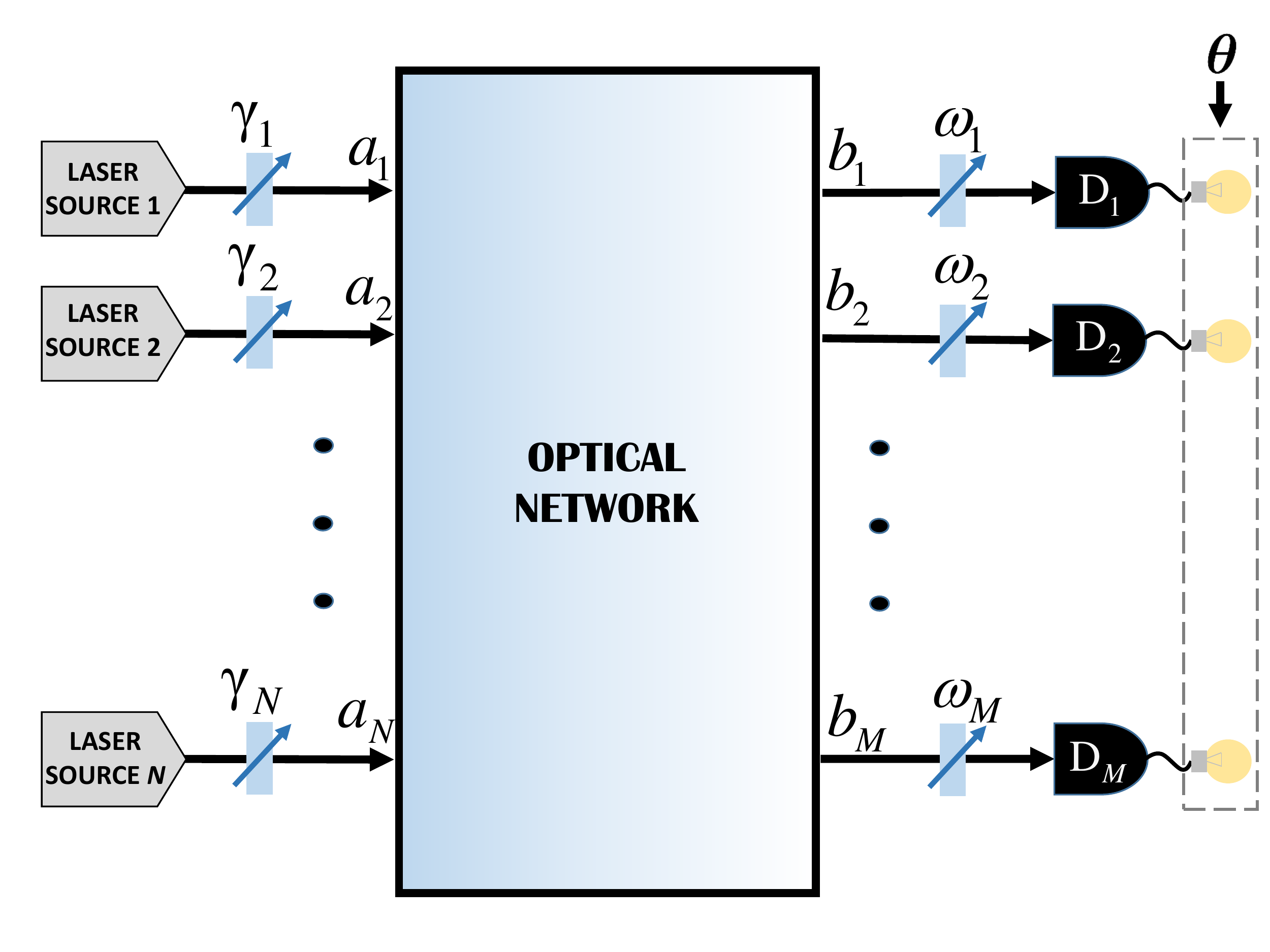}
	\end{center}
	\caption{Schematic of the method to characterize the quantum interference behaviour of an optical network (ON) by means of simple laser sources and threshold single-photon detectors. It builds on the decoy-state method~\cite{hwang2003,lo2005,wang2005a} and the detector-decoy technique~\cite{moroder2009,lim2015random}. More precisely, we place at each input port $i=1,...,N,$ of the ON a source of phase-randomized WCPs together with a variable attenuator of transmittance $\gamma_i$. At each output port $j=1,...,M$ of the ON we place a variable attenuator of transmittance $\omega_j$ and a threshold single-photon detector D$_j$. The input (output) spatial modes are denoted in the figure with the letters $a_i$ ($b_j$), and the output detection pattern of click and no-click events given by the threshold single-photon detectors is denoted by $\bm{\theta}$.
		\label{EsquemaGeneral}}
\end{figure}

More precisely, we use as input signals to the ON Fock diagonal states with different photon-number statistics. This type of signals could be generated, for instance, with attenuated laser diodes emitting phase-randomised weak coherent pulses (WCPs), triggered spontaneous parametric down-conversion sources or practical single-photon sources, together with variable attenuators to vary the intensity of the different light pulses. To implement the detector-decoy method, on the other hand, we place variable attenuators also on the output ports of the ON together with threshold single-photon detectors. This general scenario is illustrated in Fig.~\ref{EsquemaGeneral}. 
In so doing, as we show below, we have that the probability of each possible detection pattern observed on the threshold single-photon detectors can be written as a sum of linear terms where the only unknowns are the probabilities $P(x_1,...,x_M|n_1,...,n_N)\equiv P(\bf{x}|\bf{n})$ (where, for easy of notation, we use ${\bf x}\equiv x_1,...,x_M$ and ${\bf n}\equiv n_1,...,n_M$ in what follows). As a result, we obtain a set of linear equations which are function of the probabilities $P(\bf{x}|\bf{n})$ and, in principle, one can estimate these quantities accurately. The more decoy-state/detector-decoy settings we use, the higher the number of linear equations that we obtain and, thus, the better the accuracy of the estimation. Indeed, in the asymptotic limit where one uses an infinite number of decoy-state/detector-decoy settings then the probabilities $P(\bf{x}|\bf{n})$ could be estimated precisely. Importantly, however, as we show below, already a small number of decoy-state/detector-decoy settings can typically provide a quite tight estimation of $P(\bf{x}|\bf{n})$ for small values of ${\bf n}$ and ${\bf x}$. 

Our starting point is the input state to the ON. As shown in Fig.~\ref{EsquemaGeneral}, this is the state of the $N$ spatial modes after the input attenuators of transmittance $\gamma_i$. This state can be written as
\begin{eqnarray}
\rho_{\rm{in}}^{\mu}&=&\bigotimes_{i=1}^N\rho_{i}^{\mu_i}=\sum_{\bf{n}}P_{\bf{n}}^{\mu}\ket{\bf{n}}\bra{\bf{n}},
\end{eqnarray}
where $\rho_{i}^{\mu_i}=\sum_{n_i=0}^{\infty}p_{n_i}^{\mu_i}\ket{n_i}\bra{n_i}$ is the Fock diagonal state at the $i$th input spatial mode of the ON, which in the case of phase-randomized WCPs satisfies $p_{n_i}^{\mu_i}=e^{-\mu_{i}}\mu_{i}^{n_i}/n_i!$. Here, the mean photon number $\mu_i=\gamma_i \mu'$, with $\mu'$ being the initial intensity of the laser sources. The quantity $P_{{\bf n}}^{\mu}=\prod_{i=1}^N p_{n_i}^{\mu_i}$, on the other hand, represents the conditional probability of having the input state $\ket{\bf{n}}\equiv\ket{n_1,...,n_N}$ given the set of input intensities $\mu=\{\mu_1,...,\mu_N\}$.

Let us now consider the output state $\rho_{\rm{out}}^{\mu}=U\rho_{\rm{in}}^{\mu}U^{\dagger}$ of the ON, where $U$ denotes the evolution unitary operator applied by the network. We can write this state in terms of the probabilities $P(\bf{x}|\bf{n})$. For this, for convenience, we first combine the effect of each output attenuator $\omega_j$ with the detection efficiency of each threshold single-photon detector D$_j$ (see Fig.~\ref{EsquemaGeneral}). By doing so, we can conceptually consider that at the $j$th output port of the ON there is now a threshold single-photon detector with efficiency $\kappa_j=\omega_j \eta_{\rm D}$, for $j=1,...,M$, where $\eta_{\rm D}$ is the detection efficiency of the threshold single-photon detector D$_j$ in the original scenario (note that here, for simplicity, we assume that all detectors D$_j$ have the same detection efficiency $\eta_{\rm D}$). This is so because when a detector has some finite detection efficiency $\eta_{\rm D}$ it can be mathematically described by a beamsplitter of transmittance $\eta_{\rm D}$ combined with a lossless detector~\cite{yurke85}. Importantly, since the positive-operator valued measure (POVM) that characterizes the behaviour of a typical threshold single-photon detector is diagonal in the Fock bases, it follows that the resulting measurement statistics when measuring $\rho_{\rm{out}}^{\mu}$ remain unchanged if, before the actual measurements, we perform a quantum nondemolition (QND) measurement of the total number of photons at each output mode of the ON. This means, in particular, that for any $\rho_{\rm{out}}^{\mu}$, there is always a Fock-diagonal state, which we shall denote by $\tilde{\rho}_{\rm{out}}^{\mu}$, of the form
\begin{eqnarray}
\tilde{\rho}_{\rm{out}}^{\mu}&=&\sum_{\bf{x}}\bra{\bf{x}}\rho_{\rm{out}}^{\mu}\ket{\bf{x}}\ket{\bf{x}}\bra{\bf{x}}
=\sum_{\bf{x}}\bra{{\bf x}}{\it U}\rho_{\rm{in}}^{\mu}{\it U}^{\dagger}\ket{\bf{x}}\ket{\bf{x}}\bra{\bf{x}}\nonumber\\
&=&\sum_{{\bf n}}\sum_{{\bf x}}P_{{\bf n}}^{\mu}|\bra{{\bf x}}U\ket{{\bf n}}|^2\ket{{\bf x}}\bra{{\bf x}}\nonumber\\
&=&\sum_{{\bf n}}\sum_{{\bf x}\leq{\bf n}}P_{{\bf n}}^{\mu}P({\bf x}|{\bf n})\ket{{\bf x}}\bra{{\bf x}},\label{equation2}
\end{eqnarray}
that provides exactly the same measurement statistics as $\rho_{\rm{out}}^{\mu}$. In Eq.~(\ref{equation2}), $\ket{\bf{x}}\equiv\ket{x_1,...,x_M}$ is the Fock state after the QND measurements on $\rho_{\rm{out}}^{\mu}$, and $P({\bf x}|{\bf n})=|\bra{{\bf x}}U\ket{{\bf n}}|^2$ denotes the conditional probability of having such state $\ket{\bf{x}}$ given that the input state to the ON is $\ket{\bf{n}}$. Note that here, for simplicity, we consider passive networks that do not create photons, and therefore we assume that $\sum_j^M x_j\leq\sum_i^N n_i$. That is, the total number of output photons cannot be greater than the total number of input photons to the ON. This last condition is expressed in Eq.~(\ref{equation2}) with the symbol $\bf{x}\leq\bf{n}$. However, we remark that our method could be applied as well to evaluate active ONs. 

Finally, to estimate the unknown probabilities $P(\bf{x}|\bf{n})$ we need to relate them with some observable quantities. For this, we use the fact that the probability $P^{\mu,\kappa}_{\bm{\theta}}$ of observing the detection pattern $\bm{\theta} \equiv (\theta_1...\theta_M$), where $\theta_j$ is equal to zero (one) for a no-click (click) event in the threshold single-photon detector D$_j$, given the state $\tilde{\rho}_{\rm{out}}^{\mu}$ and the detectors' efficiencies $\kappa=\kappa_1,...,\kappa_M$, is given by
\begin{eqnarray}
P^{\mu,\kappa}_{{\bm \theta}}={\rm Tr}\left[\tilde\rho_{{\rm out}}^{\mu}\bigotimes_{j=1}^M \Pi_{\theta_j}^{\kappa_j}\right],\label{EqTraza}
\end{eqnarray}
with the POVM elements $\Pi_{\theta_j}^{\kappa_j}$ given by
\begin{eqnarray}
\Pi_{0}^{\kappa_j}&=&(1-p_{\rm dark})\sum_{n=0}^{\infty}(1-\kappa_{j})^{n}\ket{n}\bra{n},\nonumber\\
\Pi_{1}^{\kappa_j}&=&\mathbbm{1}-\Pi_{0}^{\kappa_j},\label{POVMset}
\end{eqnarray}
where $p_{{\rm dark}}$ denotes the dark-count probability of the detector D$_j$, which for simplicity we assume is equal for all $j=1,...,M$.  That is, the operator $\Pi_{0}^{\kappa_j}$ ($\Pi_{1}^{\kappa_j}$) is associated to a no-click (click) event at the detector D$_j$. After substituting Eqs.~(\ref{equation2}) and (\ref{POVMset}) in Eq.~(\ref{EqTraza}), we finally obtain
\begin{eqnarray}
P^{\mu,\kappa}_{\bm{\theta}}&=&\sum_{{\bf n}}\sum_{{\bf x}\leq{\bf n}}P_{{\bf n}}^{\mu}P({\bf x}|{\bf n})P^{\kappa}(\bm{\theta}|{\bf x}). \label{eqGeneral}
\end{eqnarray}
Here, $P^{\kappa}({\bm\theta}|{\bf x})=\bra{{\bf x}}\otimes_{j=1}^M \Pi_{\theta_j}^{\kappa_j}\ket{{\bf x}}$ denotes the probability of observing the detection pattern $\bm{\theta}$ given the output state $\ket{\bf{x}}$, the detection efficiencies $\kappa$ and the dark count probability $p_{\rm dark}$. If the detectors D$_j$ are well-characterized, this quantity is known. Importantly, Eq.~(\ref{eqGeneral}) relates the observed probabilities $P^{\mu,\kappa}_{\bm{\theta}}$, which can be directly measured in the actual experiment, to the unknown probabilities $P({\bf x}|{\bf n})$ via the statistics $P_{{\bf n}}^{\mu}$ and $P^{\kappa}(\bm{\theta}|{\bf x})$, which are both known a priori given the experimental parameters $\mu'$, $\eta_{\rm D}$ and $p_{\rm dark}$ together with the attenuator settings $\gamma=\{\gamma_1,...,\gamma_N\}$ and $\omega=\{\omega_1,...,\omega_M\}$. Indeed, as already mentioned previously, each decoy/detector-decoy setting provides a new linear equation which has the same unknowns $P({\bf x}|{\bf n})$ but different coefficients $P_{{\bf n}}^{\mu}$ and $P^{\kappa}(\bm{\theta}|{\bf x})$ and constant terms $P^{\mu,\kappa}_{\bm{\theta}}$. Then, by solving the set of linear equations given by Eq. (\ref{eqGeneral}) one can, in principle, estimate any conditional probability $P(\bf{x}|\bf{n})$. In what follows, we illustrate this method with two simple examples of practical interest.

\section{Evaluation}\label{SecEvaluation}
\subsection{First example: beamsplitter}
In this case, we have that the creation operators, $\hat{a}_1^{\dagger}$ and $\hat{a}_2^{\dagger}$,
for the input modes of a beamsplitter and those, $\hat{b}_1^{\dagger}$ and $\hat{b}_2^{\dagger}$, for its output modes satisfy the relations $\hat{b}_1^{\dagger}=t\hat{a}_1^{\dagger}+r\hat{a}_2^{\dagger}$ and $\hat{b}_2^{\dagger}=r'\hat{a}_1^{\dagger}+t'\hat{a}_2^{\dagger}$, where the parameters $r,t,r'$and $t'$ fulfill $|t|^2+|r|^2=1$, $|t|=|t'|$, $|r|=|r'|$ and $t'r+r't=0$~\cite{ou2007multi}. That is, if the state at the input spatial modes $a_1$ and $a_2$ is say $\ket{n_1,n_2}_{a_1,a_2}$ ({\it i.e.}, it consist in $n_1$ and $n_2$ indistinguishable photons respectively), the state at the output modes $b_1$ and $b_2$ is given by the following coherent superposition of Fock states
\begin{eqnarray}
\ket{\psi_{\rm{out}}}_{b_1,b_2}&=&\sum_{i=0}^{n_1}\sum_{j=0}^{n_2}{n_1 \choose i}{n_2 \choose j}\eta^{\frac{n_2+n_1-j-i}{2}}(1-\eta)^{\frac{j+i}{2}}\nonumber\\
&\times& (-1)^{j}\sqrt{\frac{(n_2-j+i)!(n_1-i+j)!}{n_2!n_1!}}\nonumber\\
&\times&\ket{n_1-i+j,n_2-j+i}_{b_1,b_2},\label{eqLibro1}
\end{eqnarray}
where, for simplicity, we have considered the particular case in which $r=-\sqrt{1-\eta}$, $r'=-r$ and $t'=t=\sqrt{\eta}$, with $\eta$ being the transmittance of the beamsplitter. From Eq.~(\ref{eqLibro1}) one could directly theoretically calculate the probability distribution $P(x_1,x_2|n_1,n_2)=|\braket{\psi_{{\rm out}}}{x_1,x_2}|^2$ of finding, respectively, $x_1$ and $x_2$ photons at the output ports $b_1$ and $b_2$ of the beamsplitter given that there are $n_1$ and $n_2$ photons at its input ports $a_1$ and $a_2$. Importantly, according to quantum mechanics the value of this probability strongly differs from that of a {\it classical} scenario, where the photons are considered distinguishable particles which do not interfere. The HOM dip~\cite{hong1987measurement} is a well-known example of this fact. Indeed, when two photons input a $50:50$ beamsplitter through a different input port, classical mechanics predicts a probability equal to $1/2$ of finding the two photons at different output ports of the beamsplitter, while quantum mechanics predicts (for indistinguishable photons) that this probability is equal to zero. In general, this difference between the predictions of quantum and classical mechanics can be quantified by means of the visibility, which is defined as
\begin{eqnarray}\label{vis}
V_{x_1,x_2|n_1,n_2}&:=&\frac{P(x_1,x_2|n_1,n_2)_{\rm c}-P(x_1,x_2|n_1,n_2)}{P(x_1,x_2|n_1,n_2)_{\rm c}},\ \ \ \ \
\end{eqnarray}
where the subindex c denotes the classical case, {\it i.e.}, when the photons are perfectly distinguishable.

Eq.~(\ref{vis}) has been experimentally evaluated in many different experiments over the last years. For instance, in~\cite{liu2007four} and~\cite{xiang2006demonstration}, the authors obtain visibilities $V_{2,2|2,2}$ equal to 88\% for a four-photon interference scheme within an asymmetric beamsplitter and $V_{3,3|3,3}$ equal to 92\% for a six-photon interference scheme, respectively. For this, they use type-II parametric down-conversion sources to generate pairs of entangled photons and a measurement setup with four ans six threshold single-photon detectors, respectively, in combination with beamsplitters. Also, in the experiment reported in~\cite{lopes2015atomic}, the authors interfere two bosonic atoms (instead of photons) and they observe a visibility equal to about 65\%.

We now apply our method based on two sources of phase-randomized WCPs and two threshold single-photon detectors to evaluate the visibility $V_{x_1,x_2|n_1,n_2}$. Like in the general case considered in the previous section, it is straightforward to show that by varying the intensity $\mu_i$ of the input signals at the $i$-th input port of the beamsplitter as well as the attenuator's transmittance $\omega_j$ (and thus the effective detector's efficiency $\kappa_j$) at its $j$-th output port, with $j=1,2$, one can generate an arbitrary number of inequalities that involve the unknown probabilities $P(x_1,x_2|n_1,n_2)$. The final system of linear equations, particularized from Eq.~(\ref{eqGeneral}), is given by
\begin{eqnarray}
	P^{\mu,\kappa}_{\bm{\theta}}&=&\sum_{n_1,n_2}\sum_{\scriptstyle x_1,x_2\atop\scriptstyle x_1+x_2\leq n_1+n_2}P_{n_1,n_2}^{\mu}P(x_1,x_2|n_1,n_2)\nonumber\\
	&\times&P^{\kappa}(\bm{\theta}|x_1,x_2),\label{BSeq}
\end{eqnarray}
for each one of the four possible detection patterns $\bm{\theta}\equiv(\theta_1\theta_2)\in\{00,01,10,11\}$. Again, in a real experiment the probabilities $P_{n_1,n_2}^{\mu}=e^{-(\mu_1+\mu_2)}\mu_1^{n_1}\mu_2^{n_2}/(n_1!n_2!)$ and $P^{\kappa}(\bm{\theta}|x_1,x_2)=\bra{x_1,x_2}\otimes_{j=1}^2 \Pi_{\theta_j}^{\kappa_j}\ket{x_1,x_2}$, with $\Pi_{\theta_j}^{\kappa_j}$ given by Eq.~(\ref{POVMset}), are known given the experimental sets $\mu$ and $\kappa$, as well as the value of the dark count probability of the detectors, while the probabilities $P^{\mu,\kappa}_{\bm{\theta}}$ can be directly observed in the experiment, once performed. For our simulations we use as observed values $\rm{P}^{\mu,\kappa}_{\bm{\theta}}$ those predicted by quantum mechanics (see~\ref{ApPObs} for more details).

To solve the set of linear equations given by Eq.~(\ref{BSeq}) one can use analytical or numerical tools. For simplicity, here we solve Eq.~(\ref{BSeq}) numerically. For this, we first transform the set of equalities given by Eq.~(\ref{BSeq}), which contains an infinite number of unknowns $P(x_1,x_2|n_1,n_2)$, into a set of inequalities with a finite number of unknowns, as shown in~\ref{FiniteSet}. Also, we use the linear programming solver Gurobi~\cite{Gurobi} and the Matlab interface Yalmip~\cite{Yalmip}. 

\begin{figure}
  \includegraphics[width=1\columnwidth]{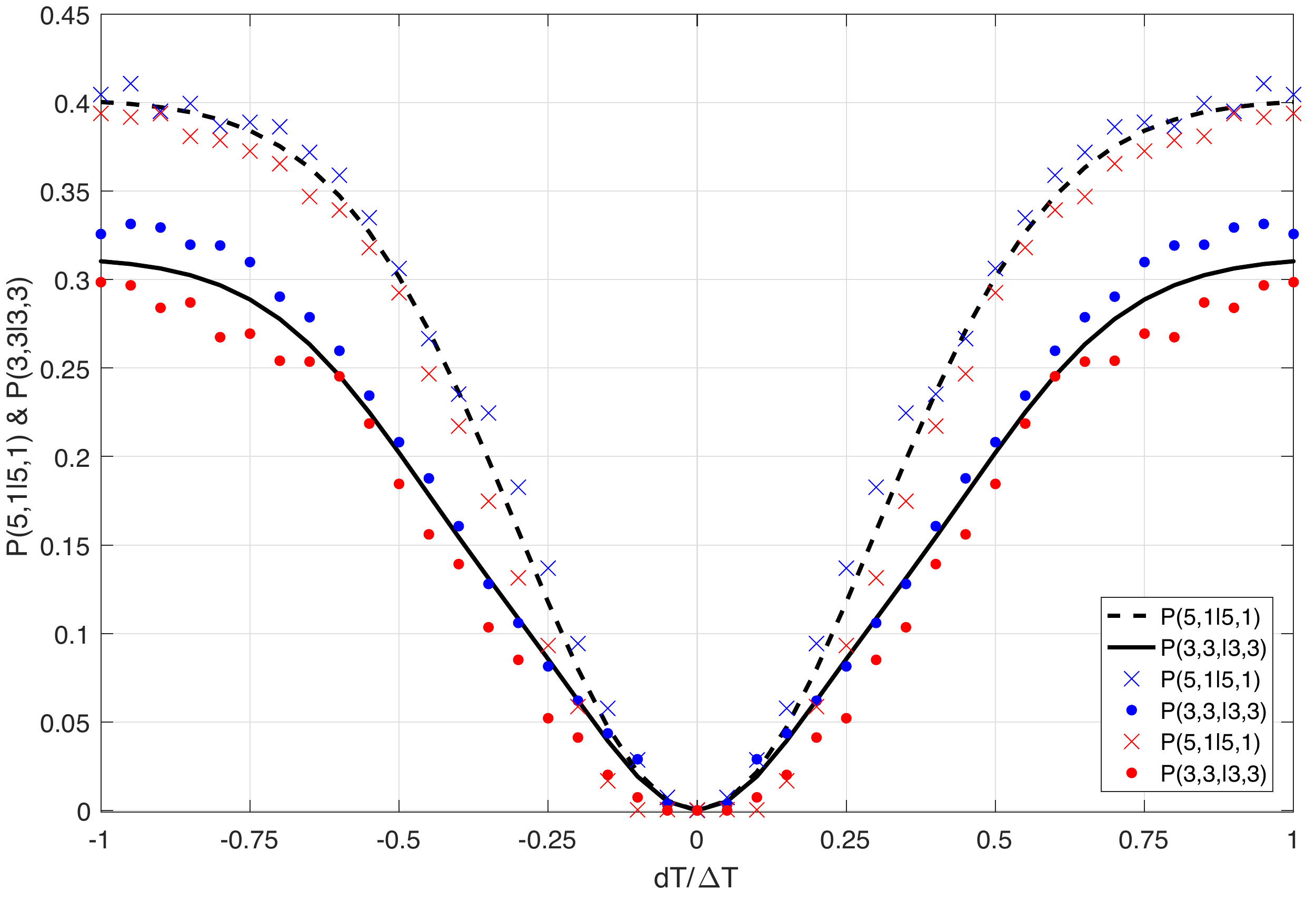}
	\caption{Hong-Ou-Mandel dip for the conditional probabilities P$(3,3|3,3)$ and P$(5,1|5,1)$ at a beamsplitter of transmittance $\eta=1/2$ and $\eta=5/6$ respectively, as a function of the relative delay $dT/\Delta T$. Here, $dT$ denotes the absolute delay between the arrival times of the optical pulses at each input port of the beamsplitter and $\Delta T$ is the FWHM of the pulses. The theoretical values predicted by quantum mechanics are illustrated with solid and dashed black lines, respectively. The blue (red) dots and crosses show the upper (lower) bound for these probabilities obtained with our method based on the use of two laser sources emitting phase-randomized WCPs and two threshold single-photon detectors. In our simulations we consider that the efficiency $\eta_{\rm D}$ of the detectors is 80\%~\cite{hadfield2009single}, the dark count probability is $p_{\rm dark}=10^{-6}$, and the transmittances $\gamma_i$ ($\omega_j$) take six (five) different values. \label{DIP}}
\end{figure}

Just as an example, Fig.~\ref{DIP} shows our results for the conditional probabilities P$(3,3|3,3)$ and P$(5,1|5,1)$ in a beamsplitter with transmittance $\eta=1/2$ and $\eta=5/6$ respectively, as a function of the relative delay $dT/\Delta T$ between the arrival times of the phase-randomized WCPs at the two input ports of the beamsplitter. Here $dT$ denotes the absolute delay between the arrival times of the optical pulses at each input port of the beamsplitter and $\Delta T$ is the full-width-half-maximum (FWHM) of the pulses, which for simplicity we assume is equal for all of them. In these simulations, the efficiency of the threshold single-photon detectors is set equal to 80\%~\cite{hadfield2009single}, and the dark count probability is $p_{\rm dark}=10^{-6}$. We have chosen these particular examples because quantum mechanics predicts that these probabilities are equal to zero ({\it i.e.}, complete destructive interference) when $dT/\Delta T=0$. As we can see from Fig.~\ref{DIP}, our estimations approximate very well the theoretical value, and the simulated lower bounds for the visibilities $V_{3,3|3,3}$ and $V_{5,1|5,1}$ are very close to one. To be precise, we obtain $V_{3,3|3,3}\geq0.99994$ and $V_{5,1|5,1}\geq0.99996$.
The reasons for the slightly noisy behaviour of the estimated values as well as for the small discrepancy between these and the theoretical values predicted by quantum mechanics (especially when $dT/\Delta T\neq0$) are mainly twofold. First, as we have already mentioned above, in our simulations we use a relatively small number of decoy-state/detector-decoy settings. In particular, for each value of $dT/\Delta T$, we choose an optimized set of six possible values for the input parameters $\mu_{1}$ and $\mu_{2}$ and five possible values for the output parameters $\kappa_{1}$ and $\kappa_{2}$. By using a larger number of settings one could in principle approximate the theoretical value as much as desired. The second reason is the limited numerical precision of the linear solver as well as the fact that, as explained in~\ref{FiniteSet}, to solve Eq.~(\ref{BSeq}) numerically we reduce the number of unknowns $P(x_1,x_2|n_1,n_2)$ to a final set. Also, we emphasise that the upper and lower bounds illustrated in Fig.~\ref{DIP} depend on the absolute value of $dT/\Delta T$. This is because the experimental data $P^{\mu,\kappa}_{\bm{\theta}}$ that we use in our simulations depend on $|dT/\Delta T|$ (see ~\ref{ApPObs} for further details).

Finally, let us remark that when we try to estimate the conditional probabilities $P(x_1,x_2|n_1,n_2)$ for higher total input photon numbers, the accuracy of the estimation decreases. This is so because the value of the coefficients $P_{{\bf n}}^{\mu}P^{\kappa}(\bm{\theta}|{\bf x})$ decreases very rapidly when $\bf n$ increases, which renders the estimation problem difficult to solve numerically even with strong scaling methods. Moreover, increasing the value of the intensity setting $\mu$ is not of much help here, since it entails an increase of the leftover term (see~\ref{FiniteSet}). Possible solutions might be to try to solve the set of linear equations analytically by means of say Gaussian elimination, or to develop more efficient numerical estimation methods. It would be definitively interesting to further investigate these two options.

\subsection{Second example: tritter}
We now estimate the three coincidence detection probability $P(1,1,1|1,1,1)$ for a tritter for two different scenarios. Both scenarios have been experimentally analysed very recently in~\cite{Messen2017distinguishability}, where the authors used heralded single-photon sources (based on spontaneous four-wave mixing in silica-on-silicon waveguides together with three threshold single-photon detectors for heralding) in combination with a measurement setup with five threshold single-photon detectors.  If we denote by $\braket{\psi_j}{\psi_k}=r_{jk}e^{i\phi_{jk}}$ the inner product between the states of the single-photons signals at the $j$th and $k$th input ports of the tritter, quantum mechanics predicts that the probability $P(1,1,1|1,1,1)$ is given by~\cite{Messen2017distinguishability,Tichy2015Sampling}
\begin{eqnarray}
P(1,1,1|1,1,1)&=&\frac{1}{9}(2+4r_{12}r_{23}r_{32}\cos(\phi)-r_{12}^2-r_{23}^2\nonumber \\
&-&r_{31}^2)\text{, }
\end{eqnarray}
where $\phi=\phi_{12}+\phi_{23}+\phi_{31}$ is the so-called collective \textit{triad} phase.   

\begin{figure}
  \includegraphics[width=0.93\columnwidth]{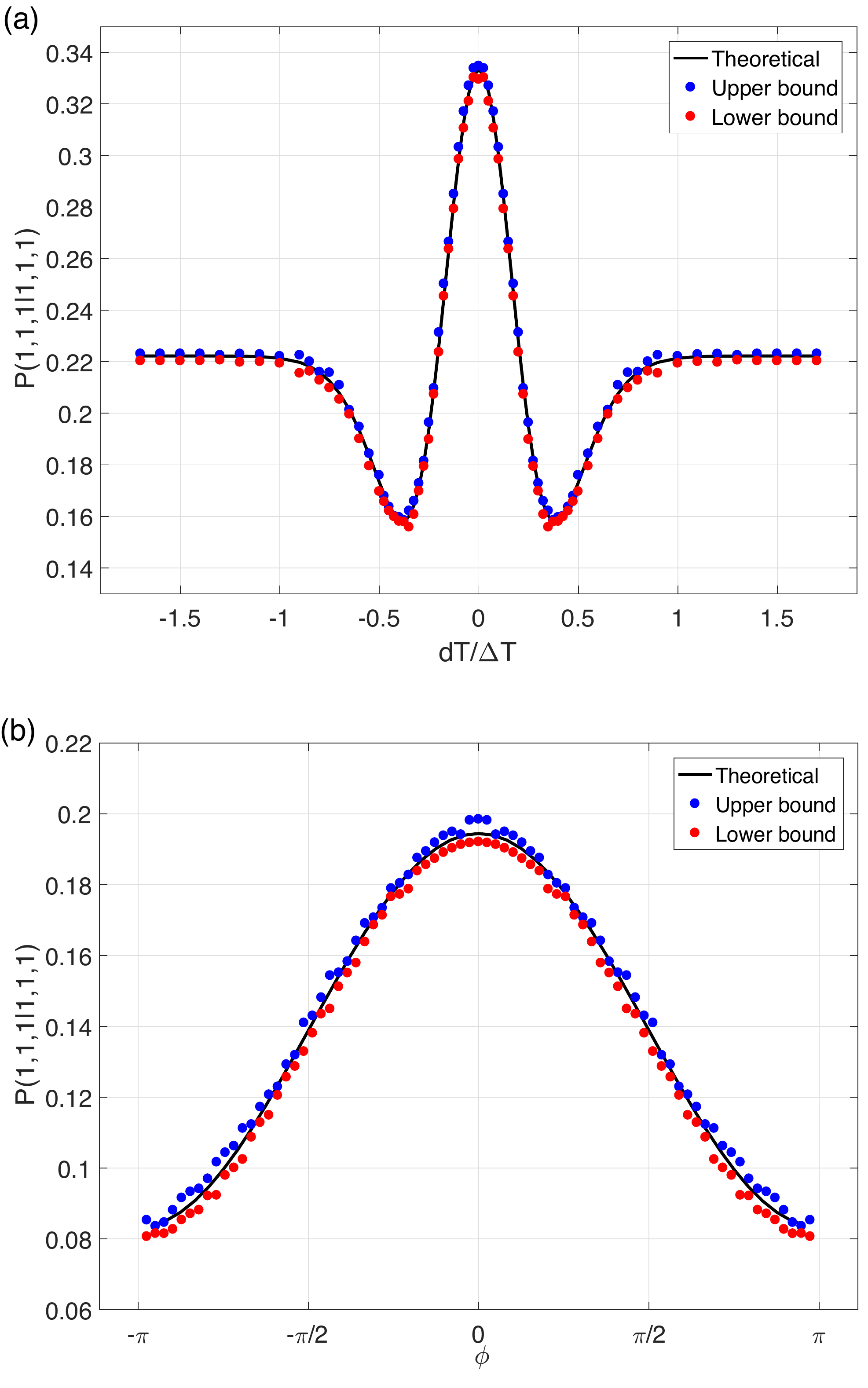}
	\caption{Three-photon coincidence probability $P(1,1,1|1,1,1)$ in a tritter. (a) Here the three input light pulses have the same polarization state and $P(1,1,1|1,1,1)$ is shown as a function of their relative delay $dT/\Delta T$. (b) In this case the three input light pulses have different polarization states and $P(1,1,1|1,1,1)$ is shown as a function of the triad phase $\phi$. In both figures, the theoretical values predicted by quantum theory are shown with a solid line while the upper and lower bounds estimated with our method are shown with dots. 
 \label{fig:Tritter}}
\end{figure}

The first scenario that we consider is shown in Fig~\ref{fig:Tritter}(a). In this case, the input pulses to the tritter have the same polarization state, but their arrival times to the different input ports of the tritter vary. The result predicted by quantum theory for $P(1,1,1|1,1,1)$ in this situation is shown with a solid line in Fig~\ref{fig:Tritter}(a), while our estimations are shown with dots. Again, we can see that the estimated upper and lower bounds for P$(1,1,1|1,1,1)$ fit tightly the theoretical probability. 
In the second scenario, now the polarization states of the input light pulses are chosen to compensate the temporal distinguishability between the arriving photons and they might be different for the signals at each input port. The motivation for this scenario is to observe the dependence that the three photon coincidence probability $P(1,1,1|1,1,1)$ has on the triad phase $\phi$ by keeping constant all those terms $r_{jk}$ that affect such probability but arise from two-photon distinguishability~\cite{Messen2017distinguishability}. The results are shown in Fig~\ref{fig:Tritter}(b), where once again we can see that our method provides a tight estimation of the theoretical values, thus showing its practicality. Also, we remark that, as in the case of Fig.~\ref{DIP}, the upper and lower bounds illustrated in Fig.~\ref{fig:Tritter} depend on $|dT/\Delta T|$ because in our simulations the experimental data $P^{\mu,\kappa}_{\bm{\theta}}$ depend on $|dT/\Delta T|$.

Moreover, like in the previous example of the beamsplitter, in our simulations we consider that the efficiency $\eta_{\rm D}$ of the threshold single-photon detectors D$_j$ is 80\%~\cite{hadfield2009single} and their dark count probability $p_{\rm dark}=10^{-6}$. Also, for the observables $P^{\mu,\kappa}_{\bm{\theta}}$ we use the expected values predicted by quantum mechanics. Furthermore, for each value of $dT/\Delta T$ in Fig.~\ref{fig:Tritter}(a), and for each value of $\phi$ in Fig.~\ref{fig:Tritter}(b), we choose three different values for the intensities of the phase-randomized WCPs as well as two possible values for the output attenuators. 

\section{Conclusion}\label{SecConclusions}
In this paper we have proposed a simple method to experimentally characterize the behaviour of small-size optical networks (ONs) on input signals that are tensor products of Fock states. More precisely, our method could be used to obtain a tight estimation of the input-output photon number statistics of a ON. Importantly, our technique could be easily implemented with current technology like, for instance, phase-randomized weak coherent pulses together with threshold single-photon detectors. The main idea of the method is rather simple: it estimates the statistics provided by ideal $n$-photon sources at the input ports of a ON by means of decoy-state techniques and it estimates the statistics provided by ideal photon-number resolving detectors at its output ports by means of detector-decoy techniques.

To illustrate the practicality of the method we have evaluated two simple examples. In the first one, we have estimated the generalized Hong-Ou-Mandel dip in a beamsplitter for a total number of six input photons, while in the second example we have estimated the three coincidence detection probability in a tritter when it receives one single-photon pulse in each of its input ports. In both cases we have obtained tight estimations that approximate very well the theoretical values.

\section{Acknowledgments}
The authors wish to thank Daniel J. Gauthier, Hoi-Kwong Lo and Norbert L\"utkenhaus for very useful discussions on the topic of this paper. This work was supported by the Galician Regional Government (consolidation of Research Units: AtlantTIC), the Spanish Ministry of Economy and Competitiveness (MINECO), the Fondo Europeo de Desarrollo Regional (FEDER) through grant TEC2014-54898-R, and the European Commission (Project QCALL). A.N. gratefully acknowledges support from a FPU scholarship from the Spanish Ministry of Education. F. X. acknowledges support from the 1000 Young Talents Program of China.

\appendix

\section{Toy model for the experimental data}\label{ApPObs}

In order to evaluate the performance of our technique we need to generate the experimental data $P^{\mu,\kappa}_{\bm{\theta}}$ which is required to run the simulations. For this, and in the absence of a real experiment, we use a simple mathematical model that we detail below. In particular, let $\bm{A}^{\dagger}$ ($\bm{B}^{\dagger}$) be the creation operators for the input (output) spatial modes of the ON. That is, $\bm{A}^{\dagger}=[\hat{a}^{\dagger}_1,...,\hat{a}^{\dagger}_N]^T$ and $\bm{B}^{\dagger}$ is defined similarly. These vectors satisfy
\begin{eqnarray}
\bm{A}^{\dagger}=U\bm{B}^{\dagger},\label{Utransf}
\end{eqnarray}
where $U$ is the unitary transformation that describes the behaviour of the ON.

In the case of WCPs, the input state to the ON can be written as $\ket{\Psi_{\text{in}}}=\bigotimes_{k=1}^N\ket{\psi_{\text{in},k}}$, where 
\begin{equation}
\ket{\psi_{\text{in},k}}=e^{\int (\alpha_{k}(\omega)\hat{a}_k^{\dagger}(\omega)-\alpha_{k}^{\ast}(\omega)\hat{a}_k(\omega))d\omega}\ket{0_k}, 
\end{equation}
\noindent is the coherent state at the $k$th input mode~\cite{blow1990continuum}. Here, the parameters $\alpha_k(\omega)$ are defined as 
\begin{equation}
\alpha_k(\omega)=\frac{\sqrt{\mu_{k}}}{(2\pi\sigma^2)^{1/4}}e^{-\frac{\omega^2}{4\sigma^2}}e^{i\phi_k-i\omega t_k}. 
\end{equation}
\noindent That is, for simplicity we assume that each $|\alpha_k(\omega)|^2$ follows a Gaussian distribution of mean zero and standard deviation $\sigma$ which is multiplied by the intensity $\mu_k$ to guarantee that the condition $\int |\alpha_k(\omega)|^2d\omega=\mu_k$ holds. The temporal parameter $t_k$ represents the arrival time of the optical pulse that enters the ON through its $k$th input port. We remark that in the definition of the states $\ket{\psi_{\text{in},k}}$ we have not included yet the fact that their phases $\phi_k$ are randomized. We will return to this point later.

Let $\{u_{jk}\}$ be the elements of the unitary matrix $U$. By applying Eq.~(\ref{Utransf}), and due to the linearity of the integral, we have that the state at the output ports of the ON can be written as $\ket{{\tilde\Psi}_{\text{out}}}=\bigotimes_{k=1}^M\ket{\psi_{\text{out},k}}$, where 
\begin{equation}
\ket{\psi_{\text{out},k}}=e^{\int (\beta_{k}(\omega)\hat{b}_k^{\dagger}(\omega)-\beta_{k}^{\ast}(\omega)\hat{b}_k(\omega))d\omega}\ket{0_k}, 
\end{equation}
\noindent and $\beta_{k}(\omega)=\sum_{j=1}^{N}\alpha_j(\omega)u_{jk}$. This means that the state $\ket{\Psi_{\text{out}}}$ at the output ports of the attenuators of efficiency $\kappa$ is given by
\begin{eqnarray}
\ket{\Psi_{\text{out}}}=e^{\sum_{k=1}^M \sqrt{\kappa_k}\int (\beta_{k}(\omega)\hat{b}_k^{\dagger}(\omega)-\beta_{k}^{\ast}(\omega)\hat{b}_k(\omega))d\omega}\ket{0}.
\end{eqnarray}
For convenience, note that here, like in the main text, we have included the effect of the efficiencies $\eta_D$ of the threshold single-photon detectors into the efficiency of the attenuators.

The probability of having vacuum in a specific output mode $k$ is related to the mean photon number $\bar{n}_k=\int |\beta_k(\omega)|^2d\omega$ of the coherent state in that mode by $\text{P}_0=e^{-\bar{n}_k}$. In order to calculate $\bar{n}_k$, let $\varphi_{jk}$ be the phase of the element $u_{jk}$ of $U$, \textit{i.e.}, $u_{jk}=|u_{jk}|e^{i\varphi_{jk}}$. Then, it is straightforward to show that
\begin{eqnarray}
|\beta_k(\omega)|^2&=&\sum_{s=1}^{N-1}\sum_{j=1}^{N-s} 2|\alpha_j(\omega)||\alpha_s(\omega)||u_{jk}||u_{sk}| \nonumber \\
&\times&\cos{(\phi_j-\phi_s+\varphi_{kj}-\varphi_{ks}+\omega (t_s-t_j))} \nonumber \\
&+&\sum_{i=1}^{N}|\alpha_i(\omega)|^2|u_{ik}|^2.
\label{betaSq}
\end{eqnarray}
This means, in particular, that 
\begin{eqnarray}
\bar{n}_k&\equiv&\int|\beta_k(\omega)|^2d\omega=\sum_{s=1}^{N-1}\sum_{j=1}^{N-s} 2\sqrt{\mu_j}\sqrt{\mu_s}|u_{jk}||u_{sk}|\nonumber \\
&\times&e^{-\frac{\tau_{js}^2 4\ln{2}}{\Delta T^2}}\cos{(\phi_j-\phi_s+\varphi_{kj}-\varphi_{ks})}\nonumber \\
&+&\sum_{j=1}^{N}\mu_j|u_{jk}|^2,\label{IntegralBeta}
\end{eqnarray}
where $\tau_{ij}=t_j-t_i$ represents the delay between the arrival times of the pulses that enter the ON through its input ports $i$th and $j$th, and $\Delta T$ is their FWHM. Finally, we have that the joint probability of detecting a certain pattern $\bm{\theta}$ on the threshold single-photon detectors is given by
\begin{eqnarray}
P^{\mu,\kappa,\bm{\phi}}_{\bm{\theta}}&=&\prod_{k=1}^{M}\left[\frac{1-(-1)^{\theta_k}}{2}+(-1)^{\theta_k}(1-p_{\rm dark})e^{-\kappa_k \bar{n}_k}\right], \nonumber \\
\end{eqnarray}
where $\bm{\phi}=\{\phi_1,...,\phi_N\}$ represents the dependence of that probability on the phase of each input coherent pulse. This is so because the probability of having no click at the output port $k$ (that is, $\theta_k=0$) is given by $P^{\mu,\kappa_k,\bm{\phi}}_{0}=(1-p_{\rm dark})e^{-\kappa_k \bar{n}_k}$, and thus the probability of having a click ($\theta_k=1$) has the form $P^{\mu,\kappa_k,\bm{\phi}}_{1}=1-(1-p_{\rm dark})e^{-\kappa_k \bar{n}_k}$.

If we consider now the fact that the input coherent states are phase-randomized, we obtain that the probability of detecting the pattern $\bm{\theta}$ on the threshold single-photon detectors D$_j$ is given by
\begin{eqnarray}
P^{\mu,\kappa}_{\bm{\theta}}=\frac{1}{(2\pi)^N}\int_{0}^{2\pi}\int_{0}^{2\pi}...\int_{0}^{2\pi}P^{\mu,\kappa,\bm{\phi}}_{\bm{\theta}}d\phi_1d\phi_2...d\phi_N, \nonumber \\
\end{eqnarray}
which can be calculated numerically or even analytically for the simplest cases.

\section{Numerical estimation with linear programming}\label{FiniteSet}

For small values of the intensities $\mu=\{\mu_1,...,\mu_N\}$ we have that the coefficients $P_{{\bf n}}^{\mu}P^{\kappa}(\bm{\theta}|{\bf x})$ of the set of linear equations given by Eq.~(\ref{eqGeneral}) drop quickly to zero when the number of photons ${\bf n}\equiv n_1,...,n_M$ increases. Therefore, one can neglect some of the terms in Eq.~(\ref{eqGeneral}) to decrease the number of unknowns $P(\bf{x}|\bf{n})$ to a finite set. For instance, one can discard all the summation terms that satisfy $\sum_i^N n_i>M_{\rm cut}$, for a certain prefixed parameter $M_{\rm cut}$. In this way, we obtain that
\begin{eqnarray}
P^{\mu,\kappa}_{\bm{\theta}}&\geq&\sum_{{\bf n}\in S_{\rm cut}}\sum_{{\bf x}\leq{\bf n}}P_{{\bf n}}^{\mu}P({\bf x}|{\bf n})P^{\kappa}(\bm{\theta}|{\bf x}),
\end{eqnarray} 
where $S_{\rm cut}$ is the subset that contains all possible {\bf n} such that $\sum_i^N n_i\leq M_{\rm cut}$. Similarly, one could also obtain an upper bound on $P^{\mu,\kappa}_{\bm{\theta}}$ that depends on the same finite number of unknowns $P({\bf x}|{\bf n})$. For this, note that
\begin{eqnarray}
P^{\mu,\kappa}_{\bm{\theta}}&=&\sum_{{\bf n}\in S_{\rm cut}}\sum_{{\bf x}\leq{\bf n}}P_{{\bf n}}^{\mu}P({\bf x}|{\bf n})P^{\kappa}(\bm{\theta}|{\bf x})\nonumber \\
&+&\sum_{{\bf n}\notin S_{\rm cut}}\sum_{{\bf x}\leq{\bf n}}P_{{\bf n}}^{\mu}P({\bf x}|{\bf n})P^{\kappa}(\bm{\theta}|{\bf x})\nonumber\\
&\leq&\sum_{{\bf n}\in S_{\rm cut}}\sum_{{\bf x}\leq{\bf n}}P_{{\bf n}}^{\mu}P({\bf x}|{\bf n})P^{\kappa}(\bm{\theta}|{\bf x}) \nonumber \\
&+&\sum_{{\bf n}\notin S_{\rm cut}}P_{{\bf n}}^{\mu}\sum_{{\bf x}\leq{\bf n}}P({\bf x}|{\bf n})\nonumber\\
&=&\sum_{{\bf n}\in S_{\rm cut}}\sum_{{\bf x}\leq{\bf n}}P_{{\bf n}}^{\mu}P({\bf x}|{\bf n})P^{\kappa}(\bm{\theta}|{\bf x})+\left[1-\sum_{{\bf n}\in S_{\rm cut}}P_{{\bf n}}^{\mu}\right]\nonumber\\
&=&\sum_{{\bf n}\in S_{\rm cut}}\sum_{{\bf x}\leq{\bf n}}P_{{\bf n}}^{\mu}P({\bf x}|{\bf n})P^{\kappa}(\bm{\theta}|{\bf x})+\Lambda_{S_{\rm cut}}^{\mu},
\end{eqnarray}
where the first inequality is due to the fact that $P^{\kappa}(\bm{\theta}|{\bf x})\leq 1$ and the second equality comes from $\sum_{{\bf x}\leq{\bf n}}P({\bf x}|{\bf n})=1$ and $\sum_{{\bf n}}P_{{\bf n}}^{\mu}=1\rm{,}$ $\forall{\bf n}$. Obviously, the leftover term $\Lambda_{S_{\rm cut}}^{\mu}=1-\sum_{{\bf n}\in S_{\rm cut}}P_{{\bf n}}^{\mu}$ should be as small as possible.
	
By using this result, one can numerically obtain an upper bound for the probability $P(\bf x|\bf n)$ by solving the following linear program
\begin{eqnarray}
\max &\textbf{  }& P(\bf x|\bf n)\nonumber\\
\text{s.t.} &&P^{\mu,\kappa}_{\bm{\theta}}\leq\sum_{{\bf n}\in S_{\rm cut}}\sum_{{\bf x}\leq{\bf n}}P_{{\bf n}}^{\mu}P({\bf x}|{\bf n})P^{\kappa}(\bm{\theta}|{\bf x})\nonumber \\
&&\quad\quad\ +\Lambda_{S_{\rm cut}}^{\mu},\textbf{  }\forall \mu,\kappa,\bm{\theta} \nonumber\\
&&P^{\mu,\kappa}_{\bm{\theta}}\geq\sum_{{\bf n}\in S_{\rm cut}}\sum_{{\bf x}\leq{\bf n}}P_{{\bf n}}^{\mu}P({\bf x}|{\bf n})P^{\kappa}(\bm{\theta}|{\bf x}),\textbf{  }\forall \mu,\kappa,\bm{\theta} \nonumber\\
&& 0\leq P({\bf x}|{\bf n})\leq 1,\textbf{  }\forall {\bf x}\leq {\bf n}, {\bf n} \in S_{\rm cut}\nonumber\\
&& \sum_{{\bf x}\leq{\bf n}}P({\bf x}|{\bf n})= 1,\textbf{  }\forall {\bf n} \in S_{\rm cut}.
\end{eqnarray}
The lower bound can be estimated by simply replacing the $\max$ with a $\min$.

\bibliographystyle{apsrev4-1}
\bibliography{bibtex_library}

\end{document}